\documentclass{elsart}

\usepackage{graphicx}

\usepackage{amssymb}

\begin{document}

\begin{frontmatter}



\title{Water Content and Superconductivity in Na$_{0.3}$CoO$_{2} \cdot y$H$_2$O}


\author[TCSAM]{J. Cmaidalka},
\author[TCSAM,Physics]{A. Baikalov},
\author[TCSAM,Physics]{Y. Y. Xue\corauthref{cor}},
\corauth[cor]{Corresponding author. Tel.: +1-713-743-8310; fax: +1-713-743-8201; e-mail: 
yxue@uh.edu}
\author[TCSAM,Physics]{R. L. Meng} and
\author[TCSAM,Physics,LBNL,HKUST]{C. W. Chu}

\address[TCSAM]{Texas Center for Superconductivity and Advanced Materials, University of Houston, 
Houston, Texas  77204-5002, USA}
\address[Physics]{Department of Physics, University of Houston, 
Houston, Texas  77204-5005, USA}
\address[LBNL]{Lawrence Berkeley National Laboratory, 1 Cyclotron Road, Berkeley, California  
94720, USA}
\address[HKUST]{Hong Kong University of Science and Technology, Hong Kong}

\begin{abstract}
We report here the correlation between the water content and superconductivity in 
Na$_{0.3}$CoO$_{2} \cdot y$H$_2$O under the influences of elevated temperature and cold 
compression. The x-ray diffraction of the sample annealed at elevated temperatures indicates 
that intergrowths exist in the compound at equilibrium when $0.6 < y < 1.4$. Its low-temperature 
diamagnetization varies linearly with $y$, but is insensitive to the intergrowth, indicative of 
quasi-2D superconductivity. The $T_c$-onset, especially, shifts only slightly with $y$. Our data 
from cold compressed samples, on the other hand, show that the water-loss non-proportionally 
suppresses the diamagnetization, which is suggestive of weak links.
\end{abstract}

\end{frontmatter}

It was recently reported \cite{tak} that the layered sodium cobalt oxyhydrate 
Na$_{x}$CoO$_{2} \cdot y$H$_2$O ($x \sim 0.3$, $y \sim 1.4$) is superconductive with a $T_c$ 
slightly below 5~K.  This compound consists of two-dimensional CoO$_2$ layers separated by 
layers of Na$^+$ and H$_2$O. The discovery has generated great interest due to the similarity 
between this cobalt oxyhydrate and cuprates. Unfortunately, our knowledge about the role of the 
H$_2$O intercalation in superconductivity is still sparse, although it seems to be one of the 
dominant factors. The $y$ was estimated by several chemical techniques as between 1.24 and 1.47 
for the fully hydrated compound \cite{tak}. Slight differences exist between two later neutron 
diffraction investigations with the lattice water contents being 1.25(2) and 1.48(3), 
respectively \cite{jor,lyn}. Several non-superconducting phases with similar structures and 
distinct water contents of $y = 0.6$, 0.3, and 0.1 have also been suggested through 
thermo-gravimetric analysis (TGA) investigation \cite{foo}.  It has been reported that the 
superconducting phase is extremely chemically unstable. Even exposure for 30~min at 
35~$^{\circ}$C in dry air may completely kill the superconductivity \cite{foo}. It has also been 
noticed that the room-temperature compression will degrade the superconductivity \cite{tak}. All 
of these degradations were attributed to the loss of water \cite{tak,foo}. Its exact nature, 
however, is not well known. Therefore, we measured the $y$, the superconducting signal size, 
and the phase composition after the samples were annealed at elevated temperature or 
cold-compressed. Our data show that: 1) intergrowth superconducting phases exist for 
$0.6 < y < 1.4$; 2) the zero-field-cooled magnetization, $M_{ZFC}$, of these intergrowth phases 
varies linearly with $y-y_0$, while the $T_c$-onset shifts negligibly, where $y_0 \sim 0.6$ is 
the water content of the non-superconducting phase; and 3) the severe suppression of $M_{ZFC}$ by 
cold compression, however, is associated with a negligible water loss and unchanged x-ray 
diffraction (XRD) patterns, suggestive of weak-links.

The parent compound Na$_{0.7}$CoO$_2$ and the de-intercalated Na$_{0.3}$CoO$_2$, as well as the 
superconducting Na$_{0.3}$CoO$_{2} \cdot y$H$_2$O phase, were synthesized following the procedure 
detailed in Refs. \cite{tak,foo,sch}. No attempt was made to adjust or measure the sodium 
content, so the reported value of 0.3 is accepted based on the weight ratio of Br:Na$_{0.7}$CoO$_2$ 
adopted in our Na de-intercalation process. The structures were determined by XRD using a Rigaku 
DMAX/BIII diffractometer.  The phase composition was estimated based upon the intensity ratio 
between the main line of each phase. The microstructure was determined with a JEOL JSM 6400 
scanning electron microscope (SEM) operated at 25~kV.  The thermo-gravimetric and differential 
thermal analysis (TGA/DTA) data were acquired using a TA Instruments SDT 2960 TGA-DTA.  
The magnetic properties were measured using a Quantum Design SQUID magnetometer.    

\begin{figure}
\begin{center}
\includegraphics[scale=1]{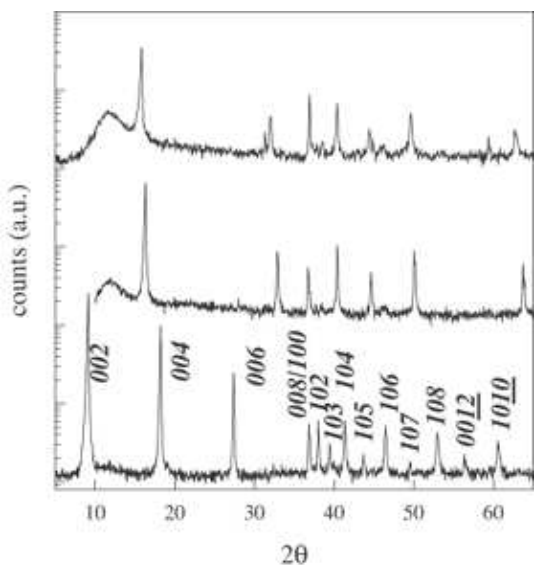}
\end{center}
\caption{XRD patterns of Na$_{0.7}$CoO$_2$ (top); Na$_{0.3}$CoO$_2$ (middle); and 
Na$_{0.3}$CoO$_{2} \cdot 1.4$H$_2$O (bottom).}
\label{fig01}
\end{figure}

\begin{figure}
\begin{center}
\includegraphics[scale=1]{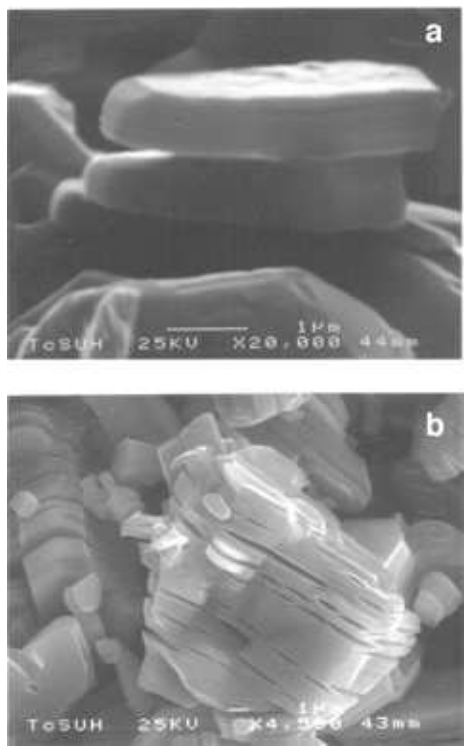}
\end{center}
\caption{SEM images of (a) Na$_{0.7}$CoO$_2$ and (b) Na$_{0.3}$CoO$_{2} \cdot 1.4$H$_2$O.}
\label{fig02}
\end{figure}

The XRD patterns are in good agreement with previous publications (Fig. \ref{fig01}). The lines in 
Fig. \ref{fig01} from top to bottom correspond to the parent compound Na$_{0.7}$CoO$_2$, the 
Na$_{0.3}$CoO$_2$ phase immediately after the Br de-intercalation, and the superconducting 
Na$_{0.3}$CoO$_{2} \cdot 1.4$H$_2$O, respectively.  All lines of 
Na$_{0.3}$CoO$_{2} \cdot 1.4$H$_2$O (bottom) above the noise floor can be indexed based on the 
$P6_{3}/mmc$ crystalline group, demonstrating that the samples are single-phase structurally. 
The SEM images of a Na$_{0.7}$CoO$_2$ sample and a Na$_{0.3}$CoO$_{2} \cdot 1.4$H$_2$O sample 
display hexagonal plate-like grains, suggesting well crystallized grains with sizes of 
1--20~$\mu$m in the $a,b$ direction (Figs. \ref{fig02}a,b). Irregular interlayer micro-cleavages, 
however, appear much more frequently in the water-intercalated samples.   

\begin{figure}
\begin{center}
\includegraphics[scale=1]{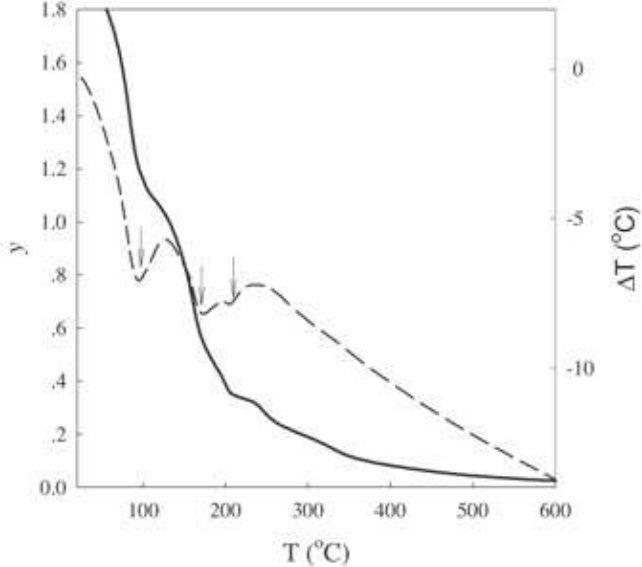}
\end{center}
\caption{TGA (solid line) and DTA (dashed line) of a fully hydrated 
Na$_{0.3}$CoO$_{2} \cdot y$H$_2$O sample with sweep rate of 20~$^{\circ}$C/min. The arrows 
show endothermic peaks.}
\label{fig03}
\end{figure}

The TGA-DTA was used to estimate the H$_2$O content under various conditions. A 
Na$_{0.3}$CoO$_{2} \cdot y$H$_2$O powder sample was measured between 25 and 600~$^{\circ}$C 
with a fast sweep rate of 20~$^{\circ}$C/min (Fig. \ref{fig03}). The apparent water content, 
$y$, was estimated based on the remaining weight above 600~$^{\circ}$C, where 
Na$_{0.3}$CoO$_{2-z}$ is assumed to be the stable phase, where $z$ represents the possible 
oxygen deficiency at elevated temperatures \cite{jor}. A $z = 0.2$ is used here to match the 
reported $y = 1.4$ at the fully hydrated state at slower sweep rate (see below). Three 
endothermic peaks can be clearly seen in the DTA trace above 50~$^{\circ}$C with 
corresponding plateaus in the TGA data. Despite the non-equilibrium nature of this fast run, 
which causes uncertainty in the values of $y$ and $T$ associated with the DTA-peaks, the 
endothermic peaks demonstrate the existence of partially dehydrated phases, as reported 
previously \cite{jor}.

\begin{figure}
\begin{center}
\includegraphics[scale=1]{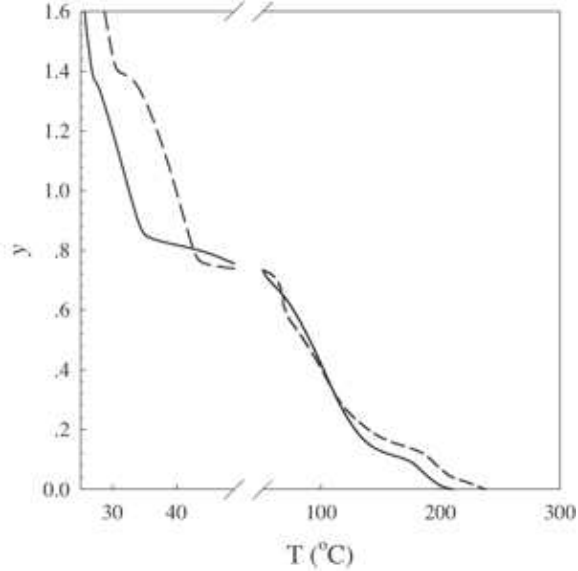}
\end{center}
\caption{TGA of Na$_{0.3}$CoO$_{2} \cdot y$H$_2$O with a sweep rate of 0.25~$^{\circ}$C/min 
under flowing oxygen (solid line) and under static room air (dashed line) of 61\% relative 
moisture at 21~$^{\circ}$C.}
\label{fig04}
\end{figure}

To further explore the equilibrium condition of Na$_{0.3}$CoO$_{2} \cdot y$H$_2$O, TGA was 
measured again with a slower sweep rate of 0.25~$^{\circ}$C/min under flowing oxygen 
(solid line in Fig. \ref{fig04}).  It is interesting to note that there is a small kink 
around 25~$^{\circ}$C with the corresponding $y = 1.4$. When TGA is done in the static room 
air (\textit{i.e.} 61\% relative moisture at 21~$^{\circ}$C), the 25~$^{\circ}$C kink 
develops into a plateau and shifts slightly to 30--34~$^{\circ}$C due to the slower 
dehydration rate, but the corresponding $y$ value remains unchanged (dashed line in 
Fig. \ref{fig04}). This is in rough agreement with the reported equilibrium condition of 
23~$^{\circ}$C and 40\% relative moisture \cite{jor}. The total water content, on the other 
hand, varies from sample to sample. The ``extra'' water above the plateau, therefore, should 
be attributed to the surface water absorbed. The heating procedure, therefore, is used to 
measure the crystalline water. The $y$ values of 0.7 and 0.1 at the respective 60~$^{\circ}$C 
and 150~$^{\circ}$C plateaus are also in rough agreement with the previous results \cite{foo}. 
The fully hydrated compound, therefore, is stable only within a rather narrow 
temperature-moisture range without extra surface water.

\begin{figure}
\begin{center}
\includegraphics[scale=1]{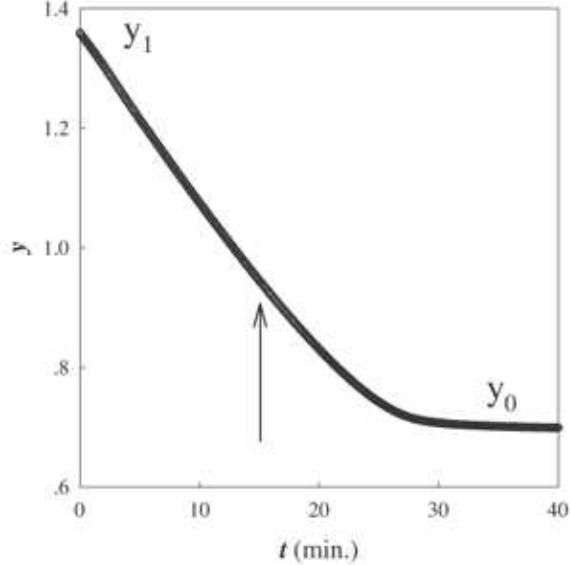}
\end{center}
\caption{Isothermal TGA curve of a Na$_{0.3}$CoO$_{2} \cdot y$H$_2$O sample at 40~$^{\circ}$C.}
\label{fig05}
\end{figure}

\begin{figure}
\begin{center}
\includegraphics[scale=1]{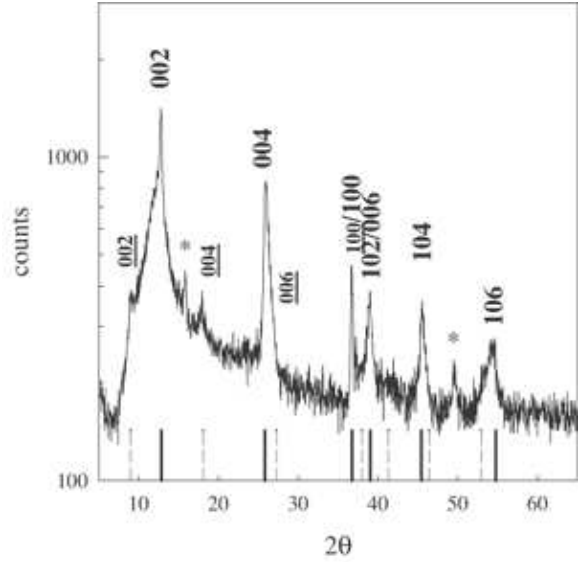}
\end{center}
\caption{The XRD of a Na$_{0.3}$CoO$_{2} \cdot 1.4$H$_2$O sample after annealing for 15~min at 
40~$^{\circ}$C: * -- Na$_{0.3}$CoO$_{2}$; solid bars with indices -- 
Na$_{0.3}$CoO$_{2} \cdot 0.6$H$_2$O; 
and dashed bars with indices underlined -- Na$_{0.3}$CoO$_{2} \cdot 1.4$H$_2$O.}
\label{fig06}
\end{figure}

Whether the phases with $0.6 < y < 1.4$ are stable and how they behave remain unknown. Jorgesen 
\textit{et al.} suggested that the Na$_{x}$CoO$_{2} \cdot y$H$_2$O should be a distinct phase with 
$x = 1/3$ and $y = 4x$ based on the structure observed \cite{jor}. The deviations from the predicted 
$1/3$ and $4/3$, in such a model, should not only affect the superconductivity, but the 
microstructure as well. The equilibrium vapor pressure reported, on the other hand, shows a 
noticeable change when $y$ varies from 1.0 to 0.6 \cite{jor}, a phenomenon not exactly consistent 
with the miscibility gap expected between two distinct equilibrium phases, but suggestive of the 
possible intermediate states. To explore this issue, TGA was acquired as a function of time at 
40~$^{\circ}$C in air (Fig. \ref{fig05}). The $y$ value dropped from $y_1 = 1.4$ to $y_0 = 0.7$ in 
30~min, which is in rough agreement with the results of Foo \textit{et al.} The XRD and 
superconductivity were then measured at both the fully hydrated state and after a 40~$^{\circ}$C, 
15~min annealing in room air, where a $y = 0.9$ is expected (the arrow in Fig. \ref{fig05}). The 
XRD pattern of the annealed sample reveals, however, a major phase of 
Na$_{0.3}$CoO$_{2} \cdot 0.6$H$_2$O with $\ll 10$\% Na$_{0.3}$CoO$_{2} \cdot 1.4$H$_2$O (dashed bars 
with indices underlined, Fig. \ref{fig06}), whose \textit{002} line can barely be noticed. The 
(\textit{002}) lines of the Na$_{0.3}$CoO$_{2} \cdot 0.6$H$_2$O (solid bars with indices, Fig. 
\ref{fig06}), 
on the other hand, are highly distorted and asymmetric. A broad shoulder on its left side spreads 
continuously to the (\textit{002}) line of Na$_{0.3}$CoO$_{2} \cdot 1.4$H$_2$O. This feature is 
inconsistent with that caused by typical microstrain and small grain size on the one hand, and 
differs from that of amorphous phases on the other hand. Intergrowth between 
Na$_{0.3}$CoO$_{2} \cdot 1.4$H$_2$O and Na$_{0.3}$CoO$_{2} \cdot 0.6$H$_2$O, therefore, is proposed, 
although the detailed structure analysis is beyond our scope. The conjecture is supported by the fact 
that the (\textit{004}) lines of the two phases are symmetric, well separated, and relatively narrow. 
The (\textit{100}) line, especially, is as sharp as that in Fig. \ref{fig01}, suggesting a rather 
long in-plane coherence length, \textit{i.e.} 100~nm or longer. The randomly distributed water 
vacancies should affect the (\textit{100}) and (\textit{00l}) lines in similar ways. This is in 
agreement with the general consideration of the lattice-distortion energy in intercalated systems. 
It should also be pointed out that the average ``extra'' water $y-y_0 \sim 0.23$ from the TGA trace 
is far greater than the possible water content of the residual Na$_{0.3}$CoO$_{2} \cdot 1.4$H$_2$O 
($\le 0.1$). Most of the ``extra'' water has to exist in the intergrowth phases. As a result, a large 
part of the CoO$_2$ layers may possess interlayer environments different from that of 
Na$_{0.3}$CoO$_{2} \cdot 1.4$H$_2$O. To verify whether these intermediate intergrowth phases are 
thermodynamically stable at a fixed $y$, 100~mg of annealed sample was sealed in a 0.5~cc plastic 
container. No noticeable changes in the XRD pattern were observed after a 12~hr, 40~$^{\circ}$C 
annealing. It should be noted that the TGA data in Fig. \ref{fig05} suggest the water-loss (intake) 
rate is relatively fast, with a time-scale of 10~min. These data, therefore, suggest that the 
intergrowth phases are stable, when the water content is fixed, and that the water off-stoichiometry may 
be realized through intergrowth.

\begin{figure}
\begin{center}
\includegraphics[scale=1]{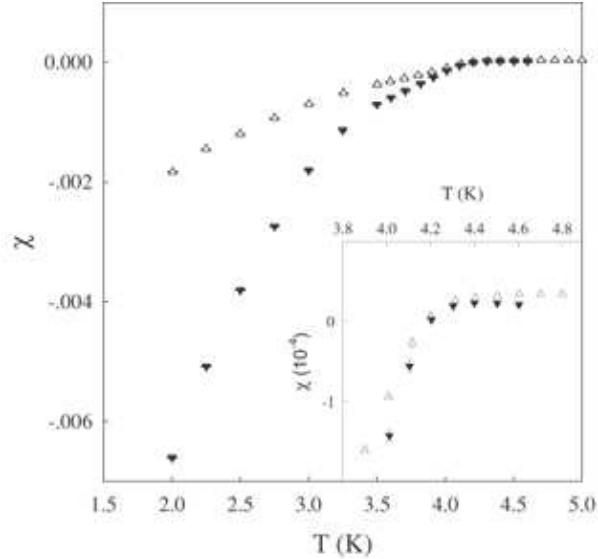}
\end{center}
\caption{The zero-field-cooled susceptibility: $\blacktriangledown$ -- 
Na$_{0.3}$CoO$_{2} \cdot 1.4$H$_2$O; $\triangle$ -- after 
15~min, 40~$^{\circ}$C annealing in air. Inset: for an expanded $T$-scale near the $T_c$-onset.}
\label{fig07}
\end{figure}

The zero-field-cooled magnetization, $M_{ZFC}$, was then measured to explore the $y$-effects on 
superconductivity (Fig. \ref{fig07}). The 15~min, 40~$^{\circ}$C annealing suppresses $|M_{ZFC}|$ by 
3.7 and 2.5 times at 2 and 3~K, respectively. $y-y_0$, on the other hand, decreases 2.9 times. The 
$|M_{ZFC}|$ suppression is far smaller than the value expected, \textit{i.e.} $> 10$, if only 
Na$_{0.3}$CoO$_{2} \cdot 1.4$H$_2$O is superconducting. Most intergrowth phases, therefore, may be 
superconducting as well, although the non-equilibrium nature of $|M_{ZFC}|$ prohibits a quantitative 
comparison. It is also interesting to note that the $T_c$ onset has not been shifted within our 
experimental resolution of 0.1~K (Inset, Fig. \ref{fig07}). The suppression of $|M_{ZFC}|$, in fact, 
is smaller when the temperature approaches $T_c$. The diamagnetic signal seems to be roughly 
proportional to $y-y_0$ with no indication of $T_c$-degradation with intergrowth. The data, therefore, 
suggest a quasi-2D superconductivity. The CoO$_2$-CoO$_2$ layer-separation, $\sim 1$~nm, is comparable 
with the $c$-axis coherence length. Those intergrowth phases with a few sequential 
(0.3Na)-0.7H$_2$O-CoO$_2$-0.7H$_2$O-(0.3Na) blocks, therefore, may possess the same superconductivity 
as that of the fully hydrated Na$_{0.3}$CoO$_{2} \cdot 1.4$H$_2$O. 

\begin{figure}
\begin{center}
\includegraphics[scale=1]{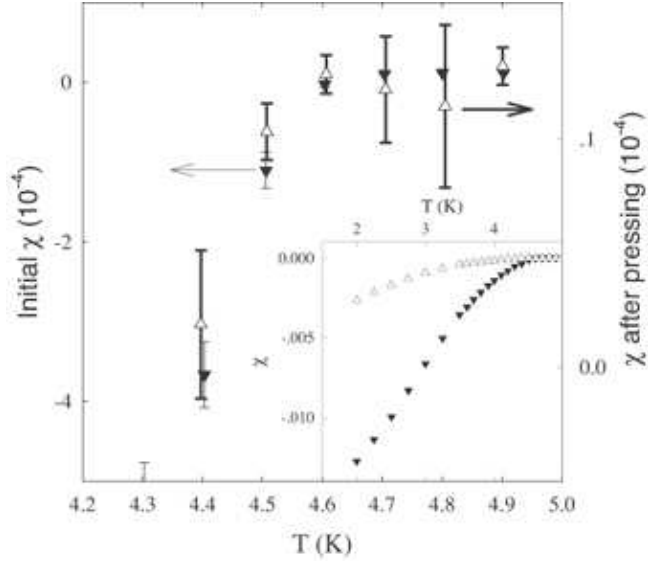}
\end{center}
\caption{The $\chi$ onset after compression. Note the different scales used. Inset: $\chi(T)$. 
$\blacktriangledown$: the initial Na$_{0.3}$CoO$_{2} \cdot 1.4$H$_2$O powder; $\triangle$: after 
11000~lb cold-compression.}
\label{fig08}
\end{figure}

\begin{figure}
\begin{center}
\includegraphics[scale=1]{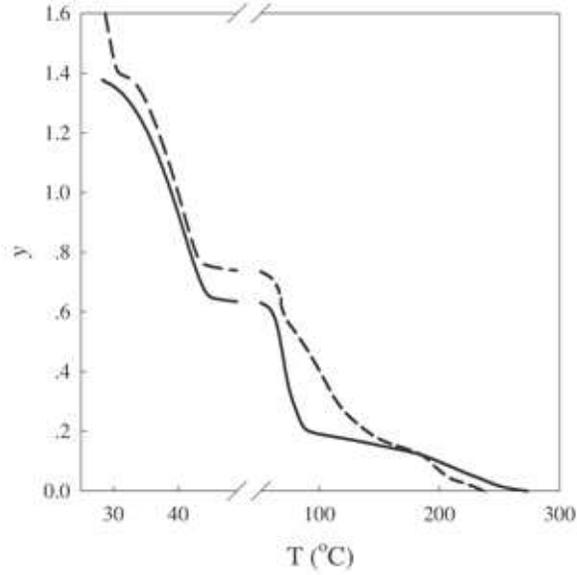}
\end{center}
\caption{TGA curves of the cold-compressed sample Na$_{0.3}$CoO$_{2} \cdot y$H$_2$O (solid line) and 
the initial powder sample of Na$_{0.3}$CoO$_{2} \cdot y$H$_2$O (dashed line).}
\label{fig09}
\end{figure}

\begin{figure}
\begin{center}
\includegraphics[scale=1]{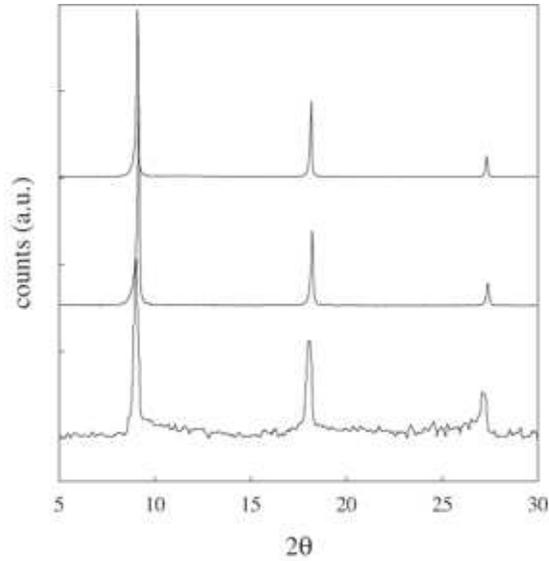}
\end{center}
\caption{The XRD from top to bottom: the initial sample; the cold-compressed sample; the ``good'' 
superconducting powder data taken from Ref. \cite{foo}.}
\label{fig10}
\end{figure}

We have also compressed several pellets at ambient temperature using a standard die-and-piston set 
(10~mm diameter).  Significant suppression of $|M_{ZFC}|$ is observed in all cases. A typical case 
with a load equal to 11000~lb is shown in the inset of Fig. \ref{fig08}. The $-\chi = -M_{ZFC}/H$ at 
2~K and 5~Oe is suppressed from the initial value ($0.17/4p$) to $0.03/4p$ by the cold compression. 
However, the weight loss due to compression is mainly the surface water and the crystalline water 
loss is rather small (Fig. \ref{fig09}). The $y$ at the 25~$^{\circ}$C plateau decreases by less than 
0.05 after the compression. This value is far from enough to cause the four-fold suppression of 
$|M_{ZFC}|$ if the $|M_{ZFC}| \propto y-y_0$ still holds. Additionally, there are no noticeable 
impurity phases in the XRD pattern, excluding the possibility of significant pressure-induced 
decomposition (Fig. \ref{fig10}).  The line-widths, which were suggested to be characteristic of 
``poor'' superconductors \cite{foo}, are the same as those of the ``good'' samples of Ref. \cite{foo}. 
In fact, the full-width at half height of the first three (\textit{00l}) lines are close to our 
instrument resolution even after the cold compression. It should also be pointed out that the 
$T_c$-onset shift is relatively small, $\sim 0.1$~K in this particular case, although shifts as large 
as 0.4~K have also been observed (Fig. \ref{fig08}). All of these observations demonstrate that the 
$T_c$ suppression under cold compression cannot be understood in terms of crystalline-water loss or 
pressure-induced decomposition alone. Weak links associated with the pressure-induced lattice defects, 
such as cracks and dislocations in grains, may be a reasonable interpretation.

In summary, we have studied the effects of water content and cold compression on the superconductivity 
of the newly discovered superconductor Na$_{0.3}$CoO$_{2} \cdot 1.4$H$_2$O.  While the compound is 
chemically unstable, the water-loss through slow annealing leads to intermediate intergrowth phases. 
The superconducting signal size is determined by the average water content, and is insensitive to the 
intergrowth. On the other hand, the severe $M_{ZFC}$ degradation after cold compression is 
accompanied only by negligible crystalline-water loss. The creation of weak links, therefore, is 
proposed to be the cause.

\section*{Acknowledgements}
The authors would like to thank Y. Y. Sun for x-ray analysis. This work is supported in part by 
NSF Grant No. DMR-9804325, the T.~L.~L.~Temple Foundation, the John J. and Rebecca Moores 
Endowment, and the State of Texas through the Texas Center for Superconductivity and 
Advanced Materials at the University of Houston; and at Lawrence Berkeley Laboratory by the 
Director, Office of Basic Energy Sciences, Division of Materials Sciences and Engineering of 
the U.S. Department of Energy under Contract No. DE-AC03-76SF00098.



\end{document}